\documentclass[a4paper, 10pt, conference]{ieeeconf}      % Use this line for a4 paper

\IEEEoverridecommandlockouts                              % This command is only needed if 
\usepackage{graphics} % for pdf, bitmapped graphics files
\usepackage{epsfig} % for postscript graphics files

\usepackage{amsmath} % assumes amsmath package installed
\usepackage{amssymb}  % assumes amsmath package installed

  \usepackage{amsthm}

\usepackage{subfigure}
\usepackage[ruled]{algorithm2e}

\newtheorem{utilityf}{Utility function}

\newtheorem{remark}{Remark}

\title{\LARGE \bf Game Theoretic Models for Profit-Sharing in Multi-fleet Platoons}

\author{Alexander Johansson and Jonas M\aa rtensson% <-this % stops a space
\thanks{This work is supported by the Strategic Vehicle Research and Innovation Programme, Horizon2020, the Swedish Foundation for Strategic Research and the Swedish Research Council.}% <-this % stops a space
\thanks{A. Johansson and J. M\aa rtensson are with the Integrated Transport Research Lab and Division of Decision and Control,
	School of Electrical Engineering and Computer Science, KTH Royal Institute
	of Technology, Stockholm, Sweden.,
	SE-100 44 Stockholm, Sweden. Emails:
	{\tt\small \{alexjoha, jonas1\}@kth.se}} %
}

\begin{document}

\maketitle
\thispagestyle{empty}
\pagestyle{empty}

%%%%%%%%%%%%%%%%%%%%%%%%%%%%%%%%%%%%%%%%%%%%%%%%%%%%%%%%%%%%%%%%%%%%%%%%%%%%%%%%
\begin{abstract}

Profit-sharing is needed within platoons in order for competing transportation companies to collaborate in forming platoons. In this paper, we propose distribution models of the profit designed for vehicles that are located at the same origin and are operated by competing transportation companies. The vehicles have default departure times, but can decide to depart at other times in order to benefit from platooning. We model the strategic interaction among vehicles with game theory and consider pure Nash equilibria as the solution concept. In a numerical evaluation we compare the outcomes of the games associated with different distribution models of the profit.

\end{abstract}

%%%%%%%%%%%%%%%%%%%%%%%%%%%%%%%%%%%%%%%%%%%%%%%%%%%%%%%%%%%%%%%%%%%%%%%%%%%%%%%%
\section{INTRODUCTION}

The transport sector emitted $24 \%$ of the total $\text{CO}_2$ emissions due to fuel combustion in 2016 and $74\%$ of the emissions from the transport sector was emitted from road transportation \cite{Publishing2018}.

Truck platooning has received attention for its ability to reduce fuel consumption of road transportation. This is demonstrated in numerical studies in \cite{Davila2013}, \cite{Bishop2017} and by field experiments in \cite{Alam2010}, \cite{Browand2004}, \cite{Tsugawa2016}, where potential fuel savings of up to $10 \%$ are reported. Truck platooning has other benefits than reduced fuel consumption, \emph{e.g.,} decreased workload and cost of drivers, improved safety by reducing the human factor and reduced traffic congestion. The interested reader is referred to \cite{Alam2015} for a high-level introduction to truck platooning.

Platoon matching is here used to denote the process of grouping vehicles (out of a pre-defined set of vehicles with fixed routes) that will form platoons. A review on planning strategies for truck platooning, including platoon matching, is given in \cite{Bhoopalam2018}. When vehicles are operated by the same transportation company, the company would seek platoon formations to maximize the total profit from platooning of all its vehicles. In contrast, when vehicles are operated by different competing transportation companies, each vehicle may instead seek its platoon formation individually to maximize its own profit from platooning.

Distributing the profit from platooning is crucial for competing transportation companies to collaborate in forming platoons. This is due to unbalance in the profit of vehicles within platoons. For example, the platoon leaders' fuel saving is small in comparison to its followers' fuel savings \cite{Browand2004}. Therefore, a vehicle needs compensation in order to agree on being a leader and contributing to the profit of its followers. The distribution model affects vehicles' individual profit given their decided platoon formations which in turn affects vehicles' decisions of platoon formations.

\begin{figure}
	\centering
	\subfigure[Vehicles are located at the origin where they are communicating and deciding on their departure times.]{\includegraphics[width=7cm]{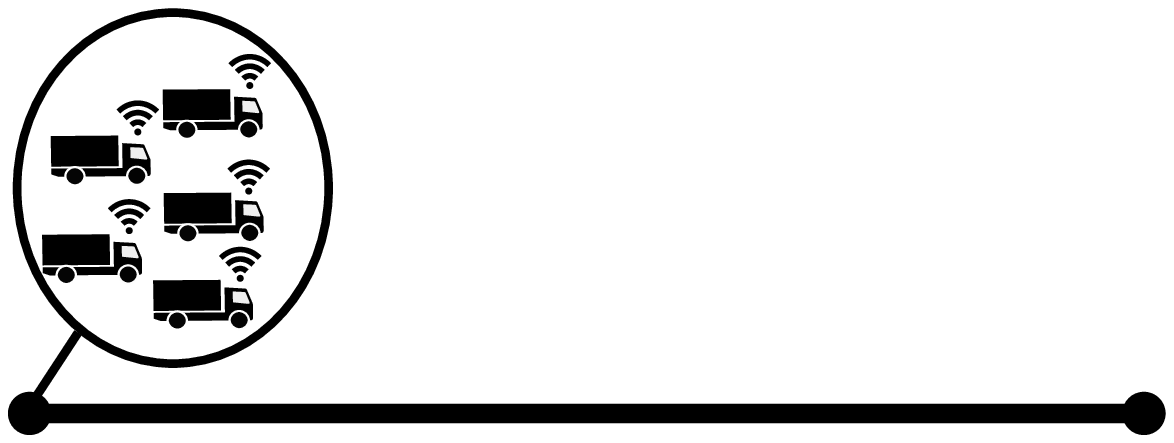}}
	\vfill
	\subfigure[Vehicles with the same departure time are forming a platoon on the route between the origin and the destination. ]{\includegraphics[width=7cm]{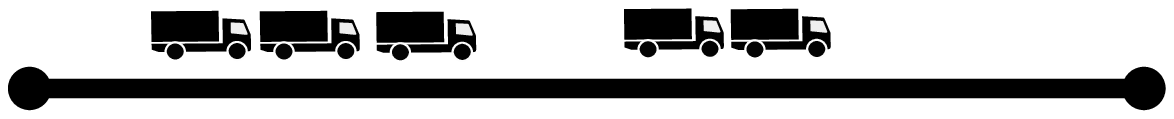}}
	\caption{Platooning scenario.}
	\label{illpark} 
\end{figure}

\subsection{Related work}

Solutions to the platoon matching problem that aim to maximize the total profit from platooning of all vehicles have been proposed in the literature, e.g., \cite{Liang2016}, \cite{Larsson2015}, \cite{Hoef2018}, \cite{Boysen2018},   \cite{Larsen2019}. 

Solutions to the platoon matching problem where vehicles individually seek to maximize their profits from platooning have been proposed in \cite{Farokhi2013} and \cite{Johansson2018}. In these solutions, each vehicle seeks to minimize its traveling cost by deciding on its departure time from a common origin. The scenario is modeled as a non-cooperative game and Nash equilibrium is used as the solution concept. We study a similar scenario as in \cite{Farokhi2013} and \cite{Johansson2018}, but our focus is to propose different distribution models of the profit and study how they affect the total profit and the platoon rate.

The authors in \cite{Sun2019} propose a solution to the platoon matching problem that aims to maximize the total profit of all vehicles. Then, the profit is distributed such that vehicles will have no incentive to leave their assigned platoons. Different to the work in \cite{Sun2019}, in this work, each vehicle seeks its platoon formation individually to maximize its own profit from platooning and the distribution model is known to the vehicles.

\subsection{Problem formulation}

We consider vehicles located at a common origin and with a common destination. The origin could be, \emph{e.g.,} a ferry terminal, a freight consolidation center or a parking place. Each vehicle has a default departure time from the origin but can decide to depart at other times in order to benefit from platooning. Vehicles with the same departure time platoon with each other from the origin to the destination. The scenario is illustrated in Figure \ref{illpark}. The vehicles are operated by different competing transportation companies and each vehicle seeks its departure time in order to maximize its own profit. The profit is described by a utility function that includes the benefit of platooning and the cost of deviating from the default departure time. The strategic interaction among vehicles is modeled as a non-cooperative game and we use Nash equilibrium as solution concept.

\begin{samepage}

The contributions of this work are:
\begin{enumerate}
	\item Four conceptual models for distributing the platooning profit among the vehicles in the platoon, denoted \textbf{even out}, \textbf{score system}, \textbf{market} and \textbf{cooperative}.
	\item A game formulation of the platoon matching problem based on the four models above and their solutions in terms of pure Nash equilibria.
	\item A numerical study of how the four models are affecting the formation of platoons, the platooning rate, and the overall profit from platooning.
\end{enumerate}

\end{samepage}

This paper is structured as follows. The platoon matching scenario that we consider is defined in Section \ref{SM}. Then, in Section \ref{MODELS}, the distribution models are proposed and a game associated with each distribution model is defined. In Section \ref{SOL}, the solution concepts for the platoon matching problems are defined and algorithms for finding the solutions are presented. The algorithms are then used when comparing the outcomes of the games associated with the distribution models in the numerical evaluation in Section \ref{NUM}. Finally, conclusions and future work are given in Section \ref{CON}.

%\begin{figure}	\label{illpark} 
%	\centering
%	\subfloat[Vehicles are located at the origin where they are communicating and deciding on their departure times.]{\includegraphics[width= 2.5in]{Illpark1}}\\	
%	\subfloat[Vehicles with the same departure time are forming a platoon on the route between the origin and the destination. ]{\includegraphics[width= 2.5in]{Illpark2}}
%	\caption{Platooning scenario.}
%	\label{illpark}
%\end{figure}

% \begin{figure}
%  \centering
%	\subfigure[Vehicles are located at the origin where they are communicating and deciding on their departure times.]{\includegraphics[width=8cm]{Illpark1}}
%\vfill
%	\subfigure[Vehicles with the same departure time are forming a platoon on the route between the origin and the destination. ]{\includegraphics[width=8cm]{Illpark2}}
%	\caption{Platooning scenario.}
	%	\label{illpark} 
%\end{figure}

\section{Platoon matching scenario}\label{SM}

Consider the scenario in Figure \ref{illpark}. Vehicles start at a common location, called origin, and have a common destination. Each vehicle has its own default departure time from the origin but can decide to depart at another time in order to benefit from platooning. The vehicles are enumerated from $1$ to $N$ and we define the index set $\mathcal N=\{1,...,N\}$.

Vehicles' decisions in the platoon matching scenario are their departure times. The departure times of vehicles are represented by integer-valued time-steps. The decided departure time of vehicle $i$ is $d_i\in \mathbb{Z}_+$. The set of departure times of the vehicles in $\mathcal N$ is denoted $\mathbf{d}=\{d_i|i\in\mathcal N\}$. The default departure time of vehicle $i$ is denoted $d_i^\star$ and the set of default departure times is denoted $\mathbf{d}^\star$. The set of departure times of all vehicles except vehicle $i$ is denoted $\mathbf{d}_{-i}=\mathbf{d}\setminus \{d_i \}$.

A vehicle can only depart at one of the default departure times in $\mathbf{d}^\star$ and each vehicle $i$ is constrained to not depart earlier than $d_i^\star$ and not later than $d_i^\star+ \delta_i$, where $\delta_i$ is the maximum delay of vehicle $i$. The set of possible departure times of vehicle $i$ is $\mathcal D_i=\{t\in\mathbf{d} | d_i^\star\le t\le d_i^\star+\delta_i\}$. The space of possible departure times of the vehicles in $\mathcal N$ is $\mathcal D=\{\boldsymbol{d}|d_i\in \mathcal D_i, \forall i\in \mathcal N  \}  $.

If two or more vehicles decide on the same departure time they will form a platoon. The index set of vehicles in the platoon that leaves at time $d_i^\star$ is 
\begin{equation*}
\mathcal N_i(\mathbf{d})=\{k\in \mathcal N | d_k=d_i^\star\}
\end{equation*}
and the number of vehicles in the platoon is the cardinality of the index set, denoted $|\mathcal N_i(\mathbf{d})|$. Note that $\mathcal N_i(\mathbf{d})$ will be empty for some $i$ when platoons are formed.

The profit from platooning differs for a platoon leader and its followers. Moreover, we allow the profit to be vehicle dependent. The profit of vehicle $i$ is $R_i^l$ when it is a platoon leader between the origin and the destination. The profit of vehicle $i$ is $R_i^f$ when it is a platoon follower between the origin and destination. Typically, $R_i^l< R_i^f$.

There is a cost associated with departing later than the default time, \emph{e.g.,} due to increased driver cost and cost related to later arrival of goods. If vehicle $i$ departs at $d_i$ its time-penalty is $B_i(d_i)$. We assume that, $B_i(d_i^\star)=0$ and $B_i(d_i)\geq 0$.

\section{Distribution models of the profit from platooning}\label{MODELS}

In this section we propose four models describing how the profit from platooning is distributed among vehicles. Each distribution model results in a model of vehicles' utility functions. Since a vehicle's profit changes whether it is a leader or a follower, a mechanism for assigning the leaders associated to each distribution model is proposed.

\subsection{Distribution model 1: even out} \label{DE}

In this model, the leader of the platoon receives a monetary compensation from its followers, according to a standardized agreement, to even out unbalance in the profit between the leader and its followers. Moreover, the platoon leader is assigned randomly.

Vehicles do not have to reveal the actual individual profit from platooning; this might be information that they want to keep secret. However, they have agreed on standard values of the profit from platooning between the origin and the destination. The standard values of the profit for being a leader and follower are denoted $R^l$ and $R^f$ respectively. In a platoon of $n>1$ members,  the transaction from each follower to the leader is $\frac{1}{n}(R^f-R^l)$. If vehicle $i$ is a leader in a platoon of $n$ vehicles, its profit from platooning after the transaction is,
\begin{equation}\label{DEeq1}
R_i^l+\frac{n-1}{n}(R^f-R^l)
\end{equation}
and if vehicle $i$ is a follower its profit is
\begin{equation}\label{DEeq2}
R_i^f -\frac{1}{n}(R^f-R^l).	
\end{equation}

\begin{remark}
	If $R^f=R_1^f=...=R_N^f$ and $R^l=R_1^l=...=R_N^l$, the profit of a leader equals the profit of its followers after the transaction. That is, \eqref{DEeq1} equals \eqref{DEeq2}.
\end{remark}

The leader in each platoon is randomly drawn from the platoon members with equal probability. In a platoon of $n$ members, the probability of each vehicle to be drawn to be a follower or the leader are $\frac{n-1}{n}$ and $\frac{1}{n}$, respectively. Then, if vehicle $i$ is in a platoon of length $n>1$, the expected profit of vehicle $i$ is
\begin{align} \nonumber
&\frac{1}{n}  \left( R_i^l+\frac{n-1}{n}(R^f-R^l) \right)+ 
\frac{n-1}{n} \left( R_i^f -\frac{1}{n}(R^f-R^l)	 \right)
\end{align}
which equals 
\begin{align} \nonumber
&\frac{1}{n} R_i^l + \frac{n-1}{n} R_i^f. 
\end{align}

\begin{remark}
	Given the random leader assignment, the expected profit from platooning after the transaction is independent of the standard profits $R_l$ and $R_f$. 
\end{remark}

\begin{utilityf}[even out] \label{modelu1}
	For each vehicle $i \in \mathcal N$ and given departure times $d_i=d_j^\star$ and $\boldsymbol d_{-i}$, if vehicle $i$ departs in a platoon with other vehicles, i.e., $|\mathcal N_{j}(d_i,\boldsymbol d_{-i})|>1$, then the utility function of vehicle $i$ is defined as
	\small
	\begin{align*}
	u_i^{eo}(d_i,\boldsymbol d_{-i})=  \!
	\frac{1}{|\mathcal N_{j}(d_i,\boldsymbol d_{-i})| }R_i^l+\frac{|\mathcal N_{j}(d_i,\boldsymbol d_{-i})|-1}{|\mathcal N_{j}(d_i,\boldsymbol d_{-i})|}R_i^f  \! -  \! B_i(d_i), 
	\end{align*}
	\normalsize where the two first terms are the expected profit of platooning and the third term is the time-penalty. If vehicle $i$ departs alone, i.e., $|\mathcal N_{j}(d_i,\boldsymbol d_{-i})|=1$, then $u_i^{eo}(d_i,\boldsymbol d_{-i})=-B_i(d_i)$.
\end{utilityf}

\subsection{Distribution model 2: score system} \label{SCORE}

Vehicles have scores that increase every time they are platoon leaders and decrease every time they are platoon followers. In each platoon formation, the vehicle with the lowest score becomes the leader. The idea is that the profit from platooning is balanced over time by the score system. 

The score of vehicle $i$ is denoted $s_i$ and the set of scores of the vehicles in $\mathcal N$ is $\boldsymbol s=\{s_i|i\in \mathcal N \}$. It is assumed that each vehicle has a unique score. If vehicle $i$ departs with other vehicles, \emph{i.e.,} $|\mathcal N_{j}(d_i,\boldsymbol d_{-i})|>1$ and it has the lowest score in the platoon, \emph{i.e.,} $s_i<s_j$ for all $j\in \mathcal N_{j}(d_i,\boldsymbol d_{-i})$, then vehicle $i$ becomes the leader. Otherwise it becomes a follower.

The scores of vehicles are updated as follows. If vehicle $i$ has score $s_i$ and becomes the leader of the platoon of vehicles in $ \mathcal N_{j}(d_i,\boldsymbol d_{-i})$, its score next time it platoons is 
\begin{equation*}
s_i+\Delta s^l\left( \mathcal N_{j}(d_i,\boldsymbol d_{-i})\right)
\end{equation*}
and if vehicle $i$ becomes a platoon follower its score next time it platoons is
\begin{equation*}
s_i-\Delta s^f\left( \mathcal N_{j}(d_i,\boldsymbol d_{-i})\right)
\end{equation*}
and if vehicle $i$ departs alone its score next time it platoons is $s_i$. Vehicle $i$ valuates each unit of score as $\beta_i$.

\begin{utilityf}[score system] \label{modelu1}
	Given departure times $d_i=d_j^\star$ and $\boldsymbol d_{-i}$, and given the scores $\boldsymbol{s}$, if vehicle $i$ departs in a platoon with other vehicles, i.e., $|\mathcal N_{j}(d_i,\boldsymbol d_{-i})|>1$, and it becomes the leader according to the score system, then its utility is
	\begin{equation*}
	u_i^s(d_i,\boldsymbol d_{-i},\boldsymbol  s)=R_i^l  -B_i(d_i)+\beta_i\Delta s^l( \mathcal N_{j}(d_i,\boldsymbol d_{-i})),
	\end{equation*}
	where the first term is its profit from platooning, the second term is the time-penalty and the third term is its valuation of the score update. If vehicle $i$ becomes a follower according to the score system its utility is 
	\begin{equation*}
	u_i^s(d_i,\boldsymbol d_{-i},\boldsymbol  s)= R_i^f  -B_i(d_i)-\beta_i\Delta s^f( \mathcal N_{j}(d_i,\boldsymbol d_{-i})).
	\end{equation*}
	If vehicle $i$ departs alone, i.e., $|\mathcal N_{j}(d_i,\boldsymbol d_{-i}))|=1$, then its utility is $u_i^s(d_i,\boldsymbol d_{-i},\boldsymbol s)=-B_i(d_i)$.

\end{utilityf}

\subsection{Distribution model 3: market} \label{AM}

A sub-set of the vehicles are sellers and the rest of the vehicles are buyers. Each seller offers the buyers to be platoon followers for a price that the seller decides. The buyers decide which seller to follow. Then, each seller seeks the price that maximizes its own profit which is a combination of its profit for being a leader and the payment it receives from the followers.

The sellers are in the set $\mathcal N_s \subseteq \mathcal N$ and the buyers are in the set $\mathcal N_b=\mathcal N \setminus \mathcal N_s $. The sellers always depart at their default departure times, more precisely, $d_i=d_i^{\star}$ for $i\in \mathcal N_s$. Moreover, sellers have unique departure times, that is, $d^\star_i\neq d_j^\star \ \forall  i,j\in \mathcal N_s$ when $i\neq j$. The departure times of all sellers are in the set $\mathcal D_s= \{d_i^\star|i\in \mathcal N_s \}$. 

The price of seller $i$ is denoted $p_i$. The set of prices of the sellers in $\mathcal N_s$ is denoted $\boldsymbol p=\{p_i|i\in \mathcal N_s \}$. The set of prices of the sellers in $\mathcal N_s$ except for seller $i$ is denoted $\boldsymbol p_{-i}=\boldsymbol p\setminus \{p_i\}$. The price of seller $i$ takes values in the finite set $\mathcal P_i$. The space of prices of the sellers in $\mathcal N_s$ is $\mathcal P = \{\boldsymbol p| p_i\in \mathcal P_i, \ \forall i \in \mathcal N_s  \}$.

Given the prices of sellers, each buyer is assumed to follow the seller that maximizes its profit, or to depart alone at its own default departure time if that is more profitable. Then, buyer $j$ decides one of the departure times in the set $\mathcal D_j^* = (\mathcal D_s \cap  \mathcal D_j) \cup \{d_j^\star \} $. If buyer $j$ departs at the default departure time of seller $i$ its profit is $R_j^f-p_i-B_j(d_i^\star)$ and if buyer $j$ departs alone at its own default departure time its profit is zero which can be written as $R_j^f-p_j-B_j(d_j^\star)$, where we define $p_j=R_j^f$ and have $B_j(d_j^{\star})=0$. Then, the most profitable departure time of buyer $j$, given the prices $\boldsymbol{p}$, is
\begin{equation}\label{buyact}  
d_j(\boldsymbol p)=
\underset{d_i^\star \in \mathcal D_j^*  }{\text{argmax}}
\ R_j^f-p_i-B_j(d_i^\star).
\end{equation}
In addition, the index set of buyers that follow seller $i$ is $F_i(p_i,\boldsymbol{p}_{-i})=\{j\in \mathcal N_b|  d_j(p_i,\boldsymbol{p}_{-i}) =d_i^\star  \} $.

\begin{remark}
	The buyers depart according to \eqref{buyact} and the profit of buyers are solely dependent on the prices of the sellers.
\end{remark}

\begin{utilityf}[market] \label{modelu3}
	For each seller $i \in \mathcal N_s$ and given prices $p_i$ and $\boldsymbol p_{-i}$, if seller $i$ is a leader for at least one buyer, i.e. $|F_i(p_i,\boldsymbol{p}_{-i})|>0$, then the utility function of vehicle $i$ is 
	\begin{equation*}
	u_i^{m}(p_i,\boldsymbol{p}_{-i} )=  R_i^l+ | F_i(p_i,\boldsymbol{p}_{-i}) |  p_i
	\end{equation*} 
	where the first term is the profit for being a leader and the second term is the received payment from  the followers. If vehicle $i$ departs without followers, i.e., $|F_i(p_i,\boldsymbol{p}_{-i} )|= 0$, then, $u_i^{m}(p_i,\boldsymbol{p}_{-i} )=0$.
	
\end{utilityf}

\begin{remark}
	The set of sellers $\mathcal N_s$ is in many cases not given. In Section \ref{SOL} an algorithm will be proposed that assigns the set of sellers.
\end{remark}

\subsection{Distribution model 4: cooperative } \label{COOP}

The distribution models \textbf{even out}, \textbf{score system} and \textbf{market} are suitable when the vehicles are operated by different competing transportation companies. A distribution model that is suitable when the vehicles are operated by the same transportation company is proposed in this sub-section. The transportation company seeks departure times of its vehicles to maximize the total profit from platooning and minimize the total time-penalty of its vehicles.

In each platoon, the leader is assigned to maximize the total profit. That is, in each platoon, the vehicle with the smallest difference in profit for being a platoon follower and platoon leader is assigned to be the platoon leader. The profit from platooning connected to vehicle $i$ is $b_i(\boldsymbol{d})=R_i^f$ if it is a follower and $b_i(\boldsymbol{d})=R_i^l$ if it is a leader. The profit connected to vehicle $i$ if it departs alone is  $b_i(\boldsymbol d)=0$.
\begin{utilityf}[cooperative] \label{modelu3}
	Given departure times  $\boldsymbol{d}$, the common utility function of the vehicles in $\mathcal N$ is
	\begin{equation*}
	u^{c}(\boldsymbol{d})= \sum \limits_{j\in \mathcal N} \big( b_j(\boldsymbol{d})-B_j(d_j) \big),
	\end{equation*}	
	which is the sum of the profit of the vehicles and their time-penalties.	
\end{utilityf}

\subsection{Spontaneous platooning} \label{SPON}

In this model, the vehicles depart at their default departure times, \emph{i.e.,} $\boldsymbol d = \boldsymbol d^\star$. Thus, the vehicles do not have any decision to make. Vehicles that have the same default departure time platoon spontaneously. The individual utility of vehicle $i$ is $u_i^{eo}(d_i^\star,\boldsymbol d_{-i}^\star)$, which is the utility function of the distribution model \textbf{even out} when vehicles depart at their default departure times.

\section{Platoon matching games and their solutions}\label{SOL}

A game associated with each distribution model is defined in Section \ref{SOL1}. The games are used to model the rational behavior of the decision-makers, and their solution decisions correspond to pure Nash equilibria of the games. The pure Nash equilibrium is defined, and an algorithm that seeks for it is proposed in Section \ref{SOL2}. In the same section, an algorithm is proposed for assigning the sellers of the game associated with the distribution model \textbf{market}.

\subsection{Defining the games}\label{SOL1}

The distribution models \textbf{even out} and \textbf{score system} result in non-cooperative games where the players are the vehicles and the decisions, decision space and utility functions are defined in Section \ref{MODELS}. The game associated with the distribution model \textbf{even out} is $G^{eo}=(\mathcal N, \mathcal D, \mathcal U^{eo}(\boldsymbol{d}))$ where $\mathcal U^{eo}(\boldsymbol d)= \{u_i^{eo}(d_i,\boldsymbol d_{-i})|i\in\mathcal N  \}$. The game associated with the distribution model \textbf{score system} is $G^s=(\mathcal N, \mathcal D, \mathcal U^s(\boldsymbol{d},\boldsymbol{s}))$ where  $\mathcal U^s(\boldsymbol d,\boldsymbol{s})= \{u_i^s(d_i,\boldsymbol d_{-i},\boldsymbol{s})|i\in \mathcal N  \}$.

The distribution model \textbf{market} results in a non-cooperative game where the players are the sellers and decisions variables are their prices. The associated game is $G^{m}=(\mathcal N_s, \mathcal P, \mathcal U^m(\boldsymbol{p}))$, where $\mathcal U^m(\boldsymbol p)= \{u_i^m(p_i,\boldsymbol{p}_{-i} )|i\in \mathcal N_s  \} $.

The distribution model \textbf{cooperative} represents a case where all vehicles are interested in maximizing a common utility function $u^{c}(\boldsymbol{d})$. It is hard to find a global maximizer to the platoon matching problem due to its combinatorial structure. Instead we seek for sub-optimal solutions by letting each vehicle $i$ seek for its departure time $d_i$ that maximizes $u^{c}(d_i,\boldsymbol{d}_{-i})$. Then, the interaction among vehicles is modeled by the  cooperative game $G^c=(\mathcal N, \mathcal D, u^{c}(\boldsymbol{d}))$ and we define $u_i^{c}(d_i,\boldsymbol{d}_{-i})=u^{c}(d_i,\boldsymbol{d}_{-i})$.

\subsection{Solutions of the games}\label{SOL2}

First, for each of the games $G^{eo}$, $G^s$ and $G^{c}$, and given the decisions $d_i$ and $\boldsymbol d_{-i}$, let $u_i(d_i,\boldsymbol d_{-i})$ temporarily denote the individual utility function of vehicle $i$. A pure Nash equilibrium is a decision profile $\boldsymbol d^* \in \mathcal D$ such that 
\begin{equation}\label{NE1}
u_i( d_i^*,\boldsymbol d_{-i}^*)\geq u_i( d_i,\boldsymbol d_{-i}^*),\ \forall d_i\in \mathcal D_i, \ \forall i\in \mathcal N. 
\end{equation}
The best-response function of vehicle $i$ given $\boldsymbol d_{-i}$ is defined as 
\begin{equation}\label{BR1}
B_i (\boldsymbol d_{-i})=  \underset{ d_i\in \mathcal D_i }{\arg\max}   \ u_i(d_i, \boldsymbol d_{-i}).
\end{equation}

\begin{remark}
	In \eqref{NE1} and \eqref{BR1}, pure Nash equilibria and the best-response function were defined for the games $G^{eo}$, $G^s$ and $G^{c}$. However, pure Nash equilibria and the best-response function can be defined similarly for the game $G^m$.  Pure Nash equilibria of the game $G^m$ are denoted $\boldsymbol p^*$ and the best-response function of seller $i$ given the prices $\boldsymbol p_{-i}$ is denoted $B_i(\boldsymbol p_{-i})$. 
\end{remark}

Algorithm \ref{alg1} seeks for pure Nash equilibria by letting each vehicle, one at a time, update its decision according to its best-response function. The limiting decision profile is a pure Nash equilibrium and it is used later in the numerical evaluation as the solution of the games.

Algorithm \ref{alg} assigns sellers of the game $G^m$ and seeks for its pure Nash equilibria. The approach is to first find an equilibrium (using Algorithm \ref{alg1}) of an initial set of sellers, possibly all vehicles. Then, one of the sellers that departs without followers becomes a buyer. Then, an equilibrium is found considering the reduced set of sellers. This procedure is repeated until all sellers depart with followers. The limiting set of sellers $\mathcal N_s^*$ is used as the set of the assigned sellers in the numerical evaluation and the solution of corresponding game is the limiting pure Nash equilibrium.

\begin{algorithm} 
	\SetKwInOut{Input}{input}\SetKwInOut{Output}{output}
	\Input{Initial decisions, $\boldsymbol d=( d_1,...,d_N)$}
	\Output{Pure Nash equilibrium, $\boldsymbol d^*=(d_1^*,..., d_N^*)  $}
	\BlankLine
	$\boldsymbol d^{old}\neq \boldsymbol d$\\	
\While{$\boldsymbol d^{old}   \neq  \boldsymbol d^{}  $ }{ $\boldsymbol d^{old}  =\boldsymbol d^{} $ \\ 
		\For{$i \in \mathcal{N}$}{
			$ d_i^{}   =B_i (\boldsymbol d_{-i} )$}
}
$\boldsymbol d^* =\boldsymbol d $
	\caption{Seeks for pure Nash equilibria.}
	\label{alg1}
\end{algorithm}

\begin{algorithm} 
	\SetKwInOut{Input}{input}\SetKwInOut{Output}{output}
	\Input{Initial sellers, $\mathcal N_s=\mathcal N$ \\
		Initial buyers, $\mathcal N_b=\emptyset$ \\
		Initial prices, $\boldsymbol p$  }
	\Output{Assigned sellers, $\mathcal N_s^*$ \\
		Pure Nash equilibrium, $\boldsymbol p^* $ 
	}

	$\mathcal A=\mathcal N$

	\While{$\mathcal A\neq\emptyset$   }{

		$\boldsymbol p^{old}\neq \boldsymbol p$\\

		\While{$\boldsymbol p^{old}   \neq  \boldsymbol p^{}  $ }{ $\boldsymbol p^{old}  =\boldsymbol p^{} $ \\ 
			\For{$i \in \mathcal{N}_s$}{
				$p_i^{}   =B_i (\boldsymbol p_{-i} )$
				
			}
		}

		$A=\{i\in \mathcal N_s||F_i|=0 \}$
		
		\If {$\mathcal A\neq\emptyset$} {	Pick one seller $i\in A$ \\	$\mathcal N_s=\mathcal N_s \cap \{i\}  $  \\
			$\mathcal N_b=\mathcal N_b \cup \{i\}  $ \\
			$\boldsymbol p=\boldsymbol p_{-i}$}
	
	}
	$\mathcal N_s^*=\mathcal N_s$ \\
	$\boldsymbol p^* =\boldsymbol p $
	\caption{Select sellers and seeks for pure Nash equilibria of the game $G^m$.}
	\label{alg}
\end{algorithm}

\section{Numerical evaluation}\label{NUM}

The proposed algorithms in Section \ref{SOL} are used here to find solutions of the platoon matching games. The games model the interaction among vehicles when different models of profit distribution are used. First, the set-up of the numerical evaluation is presented. Second, the outcomes of the platoon matching games are compared.

\subsection{Set-up of simulation}

The considered scenario is illustrated in Figure \ref{illpark} and defined in Section \ref{SM}. The distance between the origin and destination is $200$ kilometers. The benefit from platooning is, in this simulation, considered to be the reduction of fuel consumption. We assume the percentage of fuel savings of the platoon followers and leaders to be $10 \%$ and $0 \%$, respectively. Moreover, we assume that each vehicle consumes $0.35$ liters of fuel per kilometer and the fuel price is $15$ SEK (Swedish Krona) per liter. Then, the profit of the followers and leaders are  $R_1^f=...=R_N^f=105$ SEK and $R_1^l=...=R_N^l=0$, respectively.

Vehicles' default departure times from the origin are all within a $30$ minutes interval. The default departure time of each vehicle $i$ is drawn from the set $\{0,1,...,30\}$ and each outcome has equal probability, \emph{i.e.,} $ d_i^\star \sim  \mathcal U\{0,30  \}.$  The maximum time delay of vehicle $i$ is $\delta_i=10$ minutes, \emph{i.e.,} its possible departure times are in the set $\mathcal D_i=\{d_i^\star,d_i^\star+1,...,d_i^\star+10\} $.

The vehicles are penalized with $10$ SEK for each minute they deviate from their default departure time. The time-penalty for each vehicle $i$ is $B_i(d_i)=10(d_i-d_i^\star)$.

In the distribution model \textbf{score system}, in this set-up, it is assumed that the score update of the leader in a platoon equals the sum of the score updates of its followers. Then, for the platoon of the vehicles in $\mathcal N_{j}(\boldsymbol d)$ the score update of the leader is $\Delta s^l( \mathcal N_{j}(\boldsymbol d))=\frac{|\mathcal N_{j}(\boldsymbol d)|-1}{|\mathcal N_{j}(\boldsymbol d)|}$ and the score update of the followers is $\Delta s^f( \mathcal N_{j}(\boldsymbol d))=-\frac{1}{|\mathcal N_{j}(\boldsymbol d)|}$. Moreover, the vehicles valuate each score unit as $\beta_i=\frac{R^f}{4}$.

In the distribution model \textbf{market}, in this set-up, each seller decides on its price from the set $\mathcal P_i=\{\frac{1}{5}105,\frac{2}{5}105,\frac{3}{5}105,\frac{4}{5}105\}$. Note that it is not reasonable for a seller to decide a price higher or equal to $105$ SEK since then the price exceeds the profit of platooning of the buyers and the seller will then get zero followers.

\subsection{Comparison of different distribution models}

The outcomes of the games associated with the distribution models \textbf{even out}, \textbf{score system}, \textbf{market}, \textbf{cooperative} and the spontaneous platooning presented in Section \ref{SPON}, are compared. The number of vehicles $N$ is varied from $1$ to $29$ in the numerical evaluation and we keep the interval of the default departure times fixed at $30$ minutes. Monte Carlo simulations are used to approximate the expected outcome of the games. The simulation was carried out $50$ times for each game and number of vehicles $N$. 

\begin{figure}
	\centering
	\subfigure[Average individual utility.]{\includegraphics[scale=0.485]{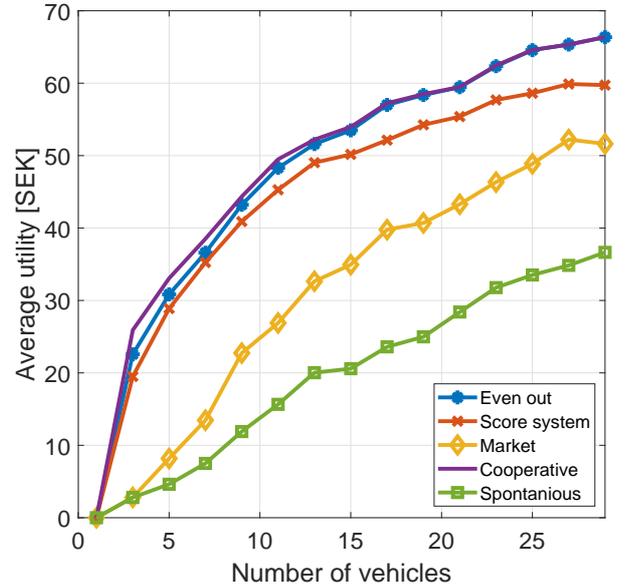}\label{COMP1}}
	\vfill
	\subfigure[Percentage of followers. ]{\includegraphics[scale=0.485]{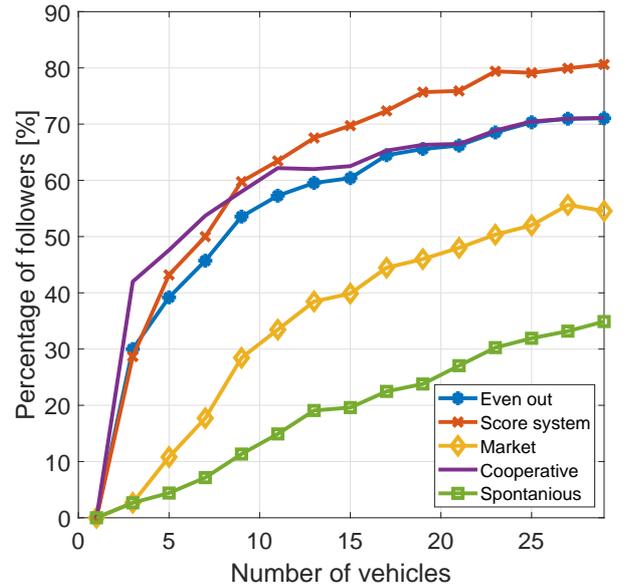}\label{COMP2}}
	\caption{Comparison of distribution models.}
	\label{case1}
\end{figure}

The average individual utility of the distribution models is shown in Figure \ref{COMP1}. We see that the highest individual utility is obtained for the \textbf{cooperative} distribution model. This is expected since the vehicles aim to maximize the total utility of all vehicles. Close to the utility of \textbf{cooperative} distribution is the utility of \textbf{even out} and the utility of \textbf{score system} is lower than the utility of the \textbf{even out}. This can be explained by the fact that vehicles with low score have low incentive to deviate from their default departure time and platooning opportunities are not exploited. The utility of \textbf{market} is low in comparison to the other distribution models. This is explained by the fact that buyers tend to spread out on sellers even when their default departure times are close which obtains lower total utility than if they depart in the same platoon. The spontaneous solution obtained lowest utility, as expected.

The percentage of followers is shown in Figure \ref{COMP2}. We see that when the number of vehicles is greater than $8$, the percentage of followers is higher in the solutions of \textbf{score system} than in \textbf{cooperative} and \textbf{even out}, even though the average utility is higher for \textbf{cooperative} and \textbf{even out}. This is possible because a higher percentage of platoon followers implies fewer platoons which can lead to higher total time-penalty and therefore lower utility.

\section{Conclusions and future work}\label{CON}

Models for distributing the profit from platooning have been proposed and the interaction among vehicles for each distribution model was modeled as a game. In a numerical evaluation it was seen that the highest individual utility was obtained when all vehicles shared both the profit from platooning and the time-penalty (\textbf{cooperative}). Moreover, it was seen that the individual utility was almost as high when the leaders received profit from its followers to even out the profit (\textbf{even out}). This suggests that, if the profit is shared in a fair way, competing companies can obtain a total profit from platooning that is near the cooperative solution, by acting selfishly. Moreover, it was seen that the spontaneous solution obtained a low utility and low platoon rate, which suggests that active platoon matching is important to increase vehicles' profit from platooning.

In future work we will extend the distribution models to be suitable for cases where the vehicles have different origins and destinations. Additionally, in our models we will capture that vehicles' profits depend on the ordering of vehicles in the platoons and design suitable profit-sharing models.

%\begin{figure}
%	\centering
%	\subfloat[][Expected average individual utility.]{\includegraphics[scale=0.44]{COMP1}\label{COMP1}}
%	\\\subfloat[][Expected percentage of followers. ]{\includegraphics[scale=0.44]{COMP2}\label{COMP2}}
%	\caption{Comparison of distribution models. }
%	\label{case1}
%\end{figure}

%\bibliographystyle{ieeetr} % 
%\bibliography{RefDatabase}

\addtolength{\textheight}{-12cm}   % This command serves to balance the column lengths

%%%%%%%%%%%%%%%%%%%%%%%%%%%%%%%%%%%%%%%%%%%%%%%%%%%%%%%%%%%%%%%%%%%%%%%%%%%%%%%%

%%%%%%%%%%%%%%%%%%%%%%%%%%%%%%%%%%%%%%%%%%%%%%%%%%%%%%%%%%%%%%%%%%%%%%%%%%%%%%%%

%%%%%%%%%%%%%%%%%%%%%%%%%%%%%%%%%%%%%%%%%%%%%%%%%%%%%%%%%%%%%%%%%%%%%%%%%%%%%%%%

\section*{ACKNOWLEDGMENT}

We thank Bj\"orn M\r{a}rdberg at Volvo Trucks for ideas, feedback and fruitful discussions.

%%%%%%%%%%%%%%%%%%%%%%%%%%%%%%%%%%%%%%%%%%%%%%%%%%%%%%%%%%%%%%%%%%%%%%%%%%%%%%%%

\bibliographystyle{ieeetr} % 
\bibliography{RefDatabase}

\end{document}